\documentstyle[12pt]{article}
\begin {document}
\baselineskip=15pt
\null\vskip -2cm
\begin{center}
{\LARGE Bosonization and Strongly Correlated Systems}

\vskip 3.0cm

{\large  Alexander O. Gogolin$^*$, Alexander A. Nersesyan$^+$}
{\large and Alexei M. Tsvelik}

\vskip 1cm

{\sl $^*$ Department of Mathematics\\
Imperial College, London}   
\vskip 1cm
{\sl $^+$ Institute of Physics, Tbilisi, Georgia}
\vskip 1cm
{\sl Department of  Physics\\
     University of Oxford\\
     Brasenose College}

\vskip 5 truecm
{\sl Cambridge University Press}\\
{\sl 1998}

\vskip 1.5cm

\end{center}
\vspace{.3cm}
\baselineskip=18pt
\vfill

%\vskip 5 truecm
%\begin{center}
%{\sl Cambridge University Press}

%{\sl 1998}
%\end{center}
\newpage

\begin{center}
{\bf Annotation}
\end{center}
\vskip 2 truecm

This volume provides a detailed account of bosonization. This important
technique represents one of the most powerful nonperturbative approaches
to many-body systems currently available.

The first part of the book examines the technical aspects of bosonization.
Topics include one-dimensional fermions, the Gaussian model, the structure
of Hilbert space in conformal theories, Bose-Einstein condensation in two
dimensions, non-Abelian bosonization, and the Ising and WZNW models.
The second part presents applications of the bosonization technique to
realistic models including the Tomonaga-Luttinger liquid, spin liquids in
one dimension and the spin-1/2 Heisenberg chain with alternative exchange.
The third part addresses the problems of quantum impurities. Chapters cover
potential scattering, the X-ray edge problem, impurities in 
Tomonaga-Luttinger liquids and the multi-channel Kondo problem.
This book will be an excellent reference for researchers and graduate
students working in theoretical physics, condensed matter physics and
field theory.

\newpage

\begin{center}
{\bf Preface}
\end{center}

\hfill {{\it We used to think that if we know one, we knew two,
because one and one are two. We are finding that we must learn a great
deal more about `and'.}\\}
%The Universe reminds me more of a big idea than of a big
%machine}\\}
\hfill{Sir Arthur Eddington, from {\it The Harvest of a Quiet Eye}, by
A. Mackay}

\hfill{{\it The behaviour of large and complex aggregations of
elementary particles, it turns out, is not to be understood in terms
of a simple extrapolation of the properties of a few
particles. Instead, at each level of complexity entirely new
properties appear, and the understanding of the new behaviours
requires research which I think is as fundamental in its nature as any
other. }

\hfill{ P. W. Anderson, from {\it More is different} (1972)}
\vskip 5mm   
High energy physics continues to fascinate people inside and outside
of science, being percieved as the `most fundamental'  area of
research. It is believed somehow that the deeper inside the matter we go
the closer we get to the truth. So it is believed that
`the truth is out there' -- at high
energies, small distances, short times. Therefore  the   
ultimate theory, Theory of
Everything, must be a theory operating at smallest distances and times possible
where there is no difference between gravitational and all other
forces (the Planck scale). All this looks extremely revolutionary and
complicated, but once  a condensed matter physicist has found  time and
courage to acquiant  himself with these ideas and theories,
these would not
appear to him
utterly unfamiliar. Moreover, despite the fact that
the two branches of physics
study   
objects of vastly  different sizes, the deeper into details you go,
the more parallels
you will find between the concepts used.
In many cases the only difference is that models
are called by different names, but this has
more to do with funding than with the essence. Sometimes differences
are more serious, but similarities still remain, for example,
the Anderson--Higgs phenomenon in particle theory is very similar
to the Meissner
effect in superconductivity; the concept of `inflation' in cosmology
is taken from the physics of first order phase transitions;
the hypothetical `cosmic strings' are similar to
magnetic field vortex lines in type II superconductors; the
Ginzburg--Landau theory of superfluid He$^3$ has many features common  
with the theory of hadron-meson interaction etc.
When you realize the existence of  this  astonishing
parallelism, it is very difficult not to think that there is something
very deep about it, that here you come across  a general principle
of Nature according to which same ideas are realized on different
space-time scales, on different hierarchical
`layers', as a Platonist would put it.
This view puts things in a new  perspective
where truth is no
longer `out there', but may be seen equally well in a `grain of
sand' as in an elementary particle.

 In this book we are going to deal with the area of theoretical
physics where the parallels between high energy and condensed matter
physics are especially strong. This area is the theory of strongly
correlated low-dimensional systems. Below we will briefly go through
these paralellisms and discuss
the history of this discipline, its main concepts, ideas and
also the features which excite interest in different
communities of physicists.

 The problems of strongly correlated systems are among the
most difficult problems of physics we are now aware of. By definition,
strongly correlated systems are those ones which cannot
be described  as a sum  of weakly interacting parts. So here we  
encounter a situation when the whole is greater than its parts, which is
always difficult to analyse.  The well-known example
of such problem in
particle physics is the problem of strong
interactions -- that is a
problem of formation and structure of heavy particles -- hadrons
(with proton and  neutron being the examples) and mesons.
In popular literature, which greatly influences minds
outside physics, one may often read
that particles constituting atomic nuclei  consist in their turn  of
`smaller', or `more elementary', particles called quarks,
coupled together with gluon fields. However, invoking images and using
language quite inadequate for  the essence of the phenomenon in
question this description more confuses than explains. The confusion
begins with the
word `consist' which here  does not have the same meaning as
when we say that a hydrogen atom consists of a proton
and an electron. This is because a hydrogen atom is formed by
electromagnetic forces and the  binding energy of the electron and proton
is small compared to their masses: $E \sim - \alpha^2 m_ec^2$, where
$\alpha = e^2/hc \approx 1/137$ is the fine structure constant and
$m_e$
is the
electron mass.
The smallness of the dimensionless coupling constant $\alpha$ obscures
the quantum character of electromagnetic  forces yielding a very small
cross section for processes of transformation of   photons into
electron--positron pairs.  Thus $\alpha$ serves as a small
parameter in a perturbation scheme where in the first approximation the
hydrogen atom is represented as a system of just two
particles. Without small $\alpha$ quantum mechanics would be a purely
academic discipline. \footnote{With only bodiless spirits to discuss it, for
sure, because there would not be stable complex atoms to form bodies.}
One cannot describe a hadron as a quantum mechanical
bound state of quarks, however,
because the corresponding fine structure constant of the strong
interactions is not small: $\alpha_G \sim 1$.
Therefore  gluon forces are of essentially quantum nature,   
in the sense that  virtual gluons
constantly emerge from vacuum and disappear, so that  the problem   
involves an infinite number of particles and therefore is
absolutely non-quantum-mechanical. It turns out, however, that the proton
and neutron have the same quantum numbers as a quantum mechanical
bound state  of three
particles of a certain kind. Only in this sense can one say that
`proton consists of three quarks'. The reader would probably agree
that this is  a very nontraditional
use  of this word.  So it is not actually a statement about
the material content of a proton (as a wave on a surface of the sea,   
it does not have
any {\it permanent}
material content), but about its symmetry  properties, that
is to what representation of the corresponding symmetry group it
belongs.

 It turns out that reduction of dimensionality may be of a great help
in solving models of strongly correlated systems. Most nonperturbative solutions
presently known (and only nonperturbative ones are  needed in physics of
strongly
correlated systems) are related to  (1 + 1)-dimensional quantum or
two-dimensional classical models. There
are two ways to relate such solutions to reality. One way is that you
imagine that reality on some level is also two dimensional. If you believe
in this you are a string theorist. Another way is to study systems where
the dimensionality is artificially
reduced. Such systems are
known in condensed matter physics; these are mostly
materials consisting of well separated chains, but there are other     
examples of effectively one-dimensional problems such as
problems of solitary magnetic impurities (Kondo effect) or of edge
states in the Quantum Hall effect. So if you are a theorist who
is interested in
seeing your predictions fulfilled during your life time,
condensed matter
physics gives you a chance.
 
 At present, there are two approaches to strongly correlated systems.
One approach, which will be only very briefly discussed in this book,   
operates with  exact solutions of many-body theories. Needless to say
not every model can be solved exactly, but fortunately
many interesting ones can. So  this
method can provide a treasury of valuable information.

 The other approach is to  try to reformulate complicated interacting   
models in such a way that they become weakly interacting. This is the
idea of bosonization which was pioneered by Jordan and Wigner in 1928
when they established equivalence between the spin $S$ = 1/2 anisotropic
Heisenberg
chain and the model of interacting fermions (we shall discuss this solution
in detail in the text). Thus in just two years after
introduction of the exclusion principle by Pauli it was
established that in  many-body  systems
the wall  separating
bosons from fermions might  become
penetrable. The example of the spin-1/2
Heisenberg chain has also  made it clear that a way to
describe  a  many-body
system is not unique, but is a matter of  convenience.

 If the anisotropy is such that the coupling
between the $z$-components of spins vanishes,  the fermionic model becomes
noninteracting. Thus, at least at this point,
the excitation spectrum (and hence  thermodynamics)
can be easely described. However, since spins are expressed in terms of the
fermionic operators in a nonlinear and  nonlocal fashion, the
problem of correlation functions remains nontrivial to the extent
that it took another
50 years to solve it.

 The transformation from spins to fermions completes the solution only
for the special value of anisotropy; at all other values fermions
interact. Interacting fermions in (1 + 1)-dimensions behave very
differently from noninteracting ones.  It turns out however, that in
many cases interactions can be effectively removed by the second
transformation -- in the given case from the fermions to a scalar
massless bosonic field. This transformation is called bosonization
and holds in the continuous
limit, that is for energies much
smaller than the bandwidth. So at such energies the spin $S$ = 1/2
Heisenberg chain can be reduced to a bunch of oscillators.

  The spin $S$ = 1/2 Heisenberg chain has provided the first example of
`particles transmutation'. We use these words to describe  a situation
when low-energy excitations of a many-body
system differ drastically from the constituent particles.  Of course, there
are elementary cases  when constituent particles are not
observable at low energies, for example,
in crystalline bodies atoms do not propagate and at low energies
we observe propagating  sound waves -- phonons; in the same way
in magnetically ordered materials instead of individual
spins we see magnons etc.
These  examples, however,  are related  to the situation  where the symmetry
is spontaneously  broken,  and the spectrum of the constituent  
particles is separated from the ground state by a gap.
Despite the fact that continuous symmetry
cannot be broken spontaneously in
(1 + 1)-dimensions and therefore there is no  finite order
parameter even at $T = 0$, spectral gaps may form. This nontrivial fact,
known as dynamical mass generation,
was discovered by Vaks and Larkin in 1961.

 However, one does not need
spectral gaps to remove single electron  excitations  
since they can be suppressed  by overdamping occuring  when
$T = 0$ is a critical point. In this case propagation of a single
particle causes
a massive emission of soft critical fluctuations.
Both  scenarios will be discussed in detail in the text.

 The fact that soft critical fluctuations may play an important role
in (1 + 1)-dimensions  became clear as soon as
theorists started to work with  such systems. It also
became clear  that the conventional methods would not work.
Bychkov, Gor'kov and
Dzyaloshinskii (1966) were the first who pointed out that instabilities of
one-dimensional metals cannot be treated in a
mean-field-like approximation. They applied to such metals an improved
perturbation series summation scheme called `parquet' approximation
(see also Dzyaloshinskii and Larkin (1972)). Originally this method
was developed for meson scattering by Diatlov, Sudakov and
Ter-Martirosyan  (1957) and
Sudakov (1957).

It was found that such  instabilities
are governed by quantum interference of two competing channels of
interaction
-- the Cooper and the Peierls ones. Summing  up  
all leading logarithmic
singularities in both channels (the {\it parquet} approximation)
Dzyaloshinskii and Larkin
obtained  differential equations for the coupling constants which later
have been  identified as
Renormalization Group equations (Solyom (1979)). From the flow of the
coupling constants one can single out the leading instabilities of
the system and thus conclude about the symmetry of the ground state.
It turned out that even in the absence of a spectral gap a coherent
propagation of single electrons is blocked.
The charge--spin separation -- one of the most striking
features of one dimensional liquid of interacting electrons -- had 
already been captured by this approach.

 The problem the diagrammatic perturbation theory could not tackle
 was
that of the strong coupling limit. Since phase transition is
not an option in (1 + 1)-dimensions, it was unclear what happens when the
renormalized interaction becomes strong (the same problem arises
for the models of
quantum impurities as the Kondo problem where  similar
singularities had also been discovered by Abrikosov (1965)).
The failure of the
conventional
perturbation theory was sealed
by P. W. Anderson (1971) who demonstrated
that it originates from what he called `orthogonality catastrophy':
the fact that the ground state wave function of an electron gas
perturbed by a local potential becomes orthogonal to the
unperturbed ground state when the number of particles goes to
infinity. \footnote{Particle transmutation includes  orthogonality
catastrophy as a particular case.} That was an indication that
the problems in question cannot
be solved by a  partial summation of
perturbation series.  This does not prevent one from trying to sum the
entire series which was brilliantly achieved
by Dzyaloshinskii and Larkin (1974) for the Tomonaga--Luttinger
 model using
the Ward identities. In fact, the subsequent development
followed the spirit of this work, but the change in formalism was
almost as dramatic  as  between the systems  of Ptolemeus and
Kopernicus.

  As it almost always happens, the  breakthrough came
from a change of the point of view. When Kopernicus put the Sun
in the centre of the coordinate frame, the
immensely complicated host of epicycles was transformed into an easily
intelligeble system of concentric orbits. In a similar way a
transformation  from fermions to bosons (hence the term {\it 
bosonization}) has provided a new convenient basis and
lead to a radical simplification of the
theory of strong  
interactions in (1 + 1)-dimensions. The bosonization method was
conceived in 1975 independently by two particle and two condensed matter
physicists   -- Sidney Coleman and
Sidney  Mandelstam, and Daniel Mattis and Alan Luther respectively.
\footnote{The first example of bosonization was  considered earlier by Schotte
and Schotte (1969).}
The focal point of their approach was the property of  Dirac fermions in (1 +
1)-dimensions. They
established  that correlation functions of such fermions
can be expressed in terms of  correlation functions of a free
bosonic field. In the bosonic representation the fermion forward
scattering
became trivial which made a solution  of the Tomonaga--Luttinger model
a simple exercise.

 The new approach had been immediately applied to previously
untreatable  problems. The results by Dzyaloshinskii and Larkin were
rederived for short range interactions and generalized to
include effects of spin. It was then understood
that low-energy sector in one-dimensional metallic systems
might  be  described by a universal effective theory
later christened `Luttinger-' or `Tomonaga--Luttinger liquid'.
The microscopic   
description of such a state was obtained by Haldane (1981), the
original idea, however, was suggested by Efetov and Larkin
(1975). Many interesting applications of bosonization to realistic
quasi-one-dimensional metals had  been considered in the 1970s
by many researches.

 Another quite fascinating discovery was also made in the 1970s and
concerns particles with fractional quantum numbers. Such particles appear as
elementary excitations in a number of one dimensional systems, with
typical example being spinons in the antiferromagnetic Heisenberg chain
with half-integer spin. A detailed description of such systems will be given
in the main text; here we just present
in the main text; here we just present
 an impressionistic picture.

 Imagine
that you have a magnet and wish  to study its excitation
spectrum. You do it by flipping individual spins and looking at
propagating waves. Naturally, since the minimal change of the total spin
projection is $|\Delta S^z| = 1$ you expect that a single flip generates a
particle of
spin-1. In measurements of dynamical spin susceptibility
$\chi''(\omega, q)$ an emission of this
particle is seen as a sharp peak. This is exactly what we see in
conventional magnets with spin-1 particles beeing magnons.

 However, in
many one dimensional systems instead of a sharp peak in
$\chi''(\omega, q)$, we  see a continuum. This means that by flipping one
spin we create at least two particles with spin-1/2. Hence fractional
quantum numbers. However, excitations with fractional spin are
subject of topological restriction -- in the given example this restriction
tells us
that the particles
can be produced only in pairs. Therefore one can say that the
elementary excitations with fractional spin (spin-1/2 in the given example)
experience
`topological  confinement'.  Topological confinement puts restriction only
on the overall number of particles leaving their spectrum unchanged. Therefore
it should be distinguished
from dynamical confinement which occurs, for instance,
in a system of two coupled spin-1/2
Heisenberg chains (see Chapter 21). There the interchain exchange confines the
spinons back to form $S = 1$ magnons giving rise to  a sharp
single-magnon peak in the neutron cross section which spreads
into  the incoherent two-spinon tail at high energies.

 An important discovery of non-Abelian bosonization was made in
1983--4 by
Polyakov and Wiegmann (1983), Witten (1984), Wiegmann (1984) and Knizhnik and
Zamolodchikov (1984). The non-Abelian approach is
very convenient when there are  spin degrees of freedom in the problem.
Its application to the Kondo model  done by Affleck and Ludwig
in the series of papers (see references in Part III)
 has drastically simplified our understanding  of
the strong coupling fixed point.

 The year 1984 witnessed  another  revolution
in low-dimensional physics. In this year Belavin, Polyakov and
Zamolodchikov published their fundamental paper on conformal field
theory (CFT). CFT provides a unified approach to all
models with gapless linear spectrum in (1 + 1)-dimensions. It was
established that if the action of a (1 + 1)-dimensional theory is
quantizable, that is its action
does not contain higher time derivatives, the linearity of the
spectrum garantees that the system has an infinite dimensional
symmetry (conformal symmetry).
The
intimate relation between CFT and the conventional bosonization had became   
manifest when Dotsenko and Fateev represented the  CFT correlation
functions in terms of correlators of bosonic exponents (1984).
In the same year Cardy (1984) and Bl\"ote, Cardy and Nightingale (1984)
found the important connection
between finite size scaling effects  and conformal
invariance.

 Both non-Abelian bosonization and CFT are steps from the initial
simplicity of the bosonization approach towards complexity of the
theory of integrable systems. Despite the fact that correlation
functions can in principle be represented in terms of correlators of
bosonic exponents, the Hilbert space of such theories is not
equivalent to the Hilbert space of free bosons. In order to make use
of the bosonic representation one must exclude certain states from the
bosonic Hilbert space. It is not always convenient to do
this directly; instead
one can
calculate the correlation functions using the Ward identities.
It is the most important result of CFT that correlation functions of
critical systems obey an infinite number of the Ward identities which
have a form of differential equations. Solving these equations  one
can uniquely determine all multi-point correlation functions. This
approach is
a substitute for the Hamiltonian formalism, since  the Hamiltonian
is effectively replaced by Ward identities for correlation
functions. Conformally invariant systems being systems with
infinite number of conservation laws constitute a subclass of exactly
solvable (integrable) models.

 After many years of intensive development the theory of strongly
correlated systems became a vast and complicated area with many
distingushed researchers working in it. Different people have different
styles and different interests -- some are concerned with applications
and some with technical developments. There is certainly a gap between
those who develop new methods and those who apply them. As an example
we can mention the Ising model which has been very extensively
studied, but scarcely used in applications. Meanwhile, as it will   
be demonstrated later in the text, this is the simplest theory
among those which remain solvable outside of criticality.

This book is
an attempt to breach the  gap between mathematics of strongly
correlated systems and its applications. In our work we have been
inspired by the idea that the theory in (1 + 1)-dimensions,
though being but a small subsector of a global theory of strongly
correlated  systems, may give an insight for more important and general
problems and give the reader a better vision of `the Universe as a
great idea'. The reader will judge whether
our  attempt is successful.
 
 In conclusion we say several words about the structure and style
of this book.
The reader should keep in mind that we shall frequently and without much
discussion switch between Hamiltonian and Lagrangian formalisms.
As a consequence  the same
notations will stand for  operators in the first case and for number
(or Grassmann number) fields in the second case. Please beware of   
this and watch what formalism is used to avoid confusion. We shall
also frequently use the field theory jargon: for example, electronic densities
are often called currents. Bear in mind that the essence of things does not
depend on how  they are called, and be indulgent.

 The book contains three parts -- in the first one we discuss the method,
in the second part describe its applications to some interesting (1 +
1)-dimensional
systems,  and in the third part discuss nonlinear quantum impurities.  
There are
important topics which we do not cover; some being even very important
-- such as
applications of bosonization in more than one spatial dimension and
the boundary
conformal theory. The only reason for omitting these topics is our ignorance.

 We would also like to explain how we selected the pictures for this
 book. We consider the pictures important since they provide
 a human element to the story and give fun. Unfortunately, we could 
 not reward  everybody on whose achievements we capitalized.
 The reason is purely technical: to draw a picture you need to meet
 the person, sit down, take time and even then the result is not
 necessarily a success. So we simply put those whom A. M. T. has
 managed to draw.

\newpage
% In parallel with these developments has run a story of integrable
%systems and exact solutions.
{\bf References}

\begin{itemize}
\item
 A. A. Abrikosov, {\it Physica} {\bf 2}, 5,  (1965).
\item
 P. W. Anderson, {\it Phys. Rev. Lett.} {\bf 18}, 1049 (1967).
\item
 A. A. Belavin, A. M. Polyakov and  A. B. Zamolodchikov,
{\it Nucl. Phys.} B{\bf 241}, 333 (1984).
\item
H. W. J. Bl\"ote,J. L. Cardy and M. P. Nightingale,
{\it Phys. Rev. Lett.} {\bf 56}, 742 (1986).
\item
 Yu. A. Bychkov, L. P. Gor'kov and I. E.  Dzyaloshinskii, {\it Sov. Phys.
JETP,} {\bf 23}, 489  (1966).
\item
J. L. Cardy, {\it J. Phys.} A{\bf 17}, L385; L961 (1984).
\item
  S. Coleman, {\it  Phys. Rev. D}{\bf 11}, 2088 (1975).
\item
 I. T. Diatlov, V. V. Sudakov and K. A. Ter-Martirosian, {\it Sov. Phys.
JETP} {\bf 5}, 631 (1957).
\item
 Vl. S. Dotsenko and V. A. Fateev, {\it Nucl. Phys.} B{\bf 240}, 312 (1984).
\item
 I. E. Dzyaloshinskii and A. I. Larkin, {\it Sov. Phys. JETP} {\bf 34}, 422
(1972).
\item
 I. E. Dzyaloshinskii and A. I. Larkin, {\it Sov. Phys. JETP} {\bf 38}, 202
(1974).
\item
K. B. Efetov and A. I. Larkin, {\it Sov. Phys. JETP,} {\bf 42}, 390 (1975).
\item
F. D. M. Haldane, {\it J. Phys.} C {\bf 14}, 2585 (1981).
\item
P. Jordan and E. Wigner, {\it Z. Phys.} {\bf 47}, 631 (1928).
\item
 V. G. Knizhnik and A. B. Zamolodchikov, {\it Nucl. Phys.} B{\bf 247}, 83
(1984).
\item
A. Luther and I. Peschel, {\it Phys. Rev.} B{\bf 9}, 2911 (1974).
\item
  S. Mandelstam, {\it Phys. Rev.} D{\bf 11}, 3026 (1975).
\item
  D. Mattis, {\it J. Math. Phys.} {\bf 15}, 609 (1974).
\item

 A. M. Polyakov and P. B. Wiegmann, {\it Phys. Lett.} B{\bf 131}, 121 (1983).
\item  
K.D. Schotte and U. Schotte,
{\it Phys. Rev.} {\bf 182}, 479 (1969).
\item  
 I. Solyom, {\it Adv. Phys.} {\bf 28}, 201 (1979).
\item
  V. V. Sudakov, {\it Sov. Phys. Doklady} {\bf 1}, 662 (1957).
\item
V. Vaks and  A. I. Larkin, {\it Sov. Phys. JETP}, {\bf 40}, 282 (1961).
\item
 P. B. Wiegmann, {\it Phys. Lett.} B{\bf
141}, 217; {\it Ibid.}, {\bf 142}, 173 (1984).
\item
 E. Witten, {\it Comm. Math. Phys.} {\bf 92}, 455 (1984).

 \end{itemize}

\newpage
\begin{center}
{\bf General bibliography}
\end{center}
\begin{itemize}
\item  
  {\it Bosonization}, collection of papers ed. by M. D. Stone,
World Scientific (1993).
\item
J. Cardy, {\it Scaling and Renormalization in Statistical Physics},  Cambridge
University Press (1996).
\item
    {\it Conformal Invariance and Applications to Statistical
Mechanics}, ed. by C. Itzykson. H. Saleur and J.-B. Zuber, World
Scientific (1988).
\item
 P. Di Francesco, P. Mathieu and D. Senechal, {\it Conformal Field
Theory}, Springer (1997).
\item
  E. Fradkin, {\it Field Theories of Condensed Matter Systems},
Addison-Wesley  (1991).
\item
  J. Fuchs, {\it Affine Lie Algebras and Quantum Groups},  Cambridge
University Press (1992).
\item
 I. S. Gradstein and I. M. Ryzhik,
{\it Tables of Integrals, Series and Products}, Academic Press,
Inc. (1980).
\item
C. Itzykson and J.-M. Drouffe, {\it Statistical Field
Theory}, Cambridge University Press (1989).
\item
L.D. Landau and E.M. Lifshits, {\it Quantum Mechanics},
Pergamon Press, Oxford, (1982).
\item
    Les Houches 1988, {\it Fields, Strings and Critical
Phenomena},  Session XLIX, ed. by E. Brezin and J. Zinn-Justin, North Holland
(1990).
\item
 B. M. McCoy and T. T. Wu, {\it The two-dimensional Ising model},
Harvard University Press (1973).
\item
    V. N. Popov, {\it Functional Integrals and Collective
Excitations}, Cambridge University Press (1990).
\item
  F. A. Smirnov, {\it Form Factors in Completely Integrable Models of
Quantum Field Theory}, World Scientific (1992).
\item
  A. M. Tsvelik, {\it Quantum Field Theory in Condensed Matter
Physics},  Cambridge University Press (1995).
\item
    J. Zinn-Justin {\it Quantum Field Theory and Critical Phenomena},
second edition,
Oxford University Press (1993).

%\end{thebibliography}
\end{itemize}

\end{document}